\documentclass[pre,twocolumn,superscriptaddress,showpacs]{revtex4}

\usepackage{amsfonts}
\usepackage{amsmath}
\usepackage{amssymb}
\usepackage{epsfig}

\begin{document}

\title{Semiclassical evaluation of quantum fidelity}
\author{Ji\v{r}\'{\i} Van\'{\i}\v{c}ek}
\affiliation{Department of Physics, Harvard University, Cambridge, Massachusetts 02138}
\affiliation{Mathematical Sciences Research Institute, Berkeley, California 94720}
\author{Eric J. Heller}
\affiliation{Department of Physics, Harvard University, Cambridge, Massachusetts 02138} 
\affiliation{Department of Chemistry, Harvard University, Cambridge, Massachusetts 02138}

\pacs{05.45.Mt, 03.65.Sq, 03.65.Yz}
\keywords{uniform semiclassical approximation, fidelity decay,
  Loschmidt echo }

\begin{abstract}
We present a numerically feasible semiclassical (SC) method to evaluate quantum fidelity decay (Loschmidt echo, FD) in a classically chaotic system.  It was thought that 
such evaluation would be intractable, but instead
we show that a uniform SC expression not only is tractable but it gives remarkably
accurate numerical results for the standard map in both the Fermi-golden-rule and Lyapunov regimes.  Because it allows
Monte Carlo evaluation, the uniform expression is accurate at times when there are $10^{70}$ 
semiclassical contributions.
Remarkably, it also explicitly contains the ``building blocks'' of analytical theories of recent literature, and 
thus permits a direct test of the approximations made by other authors
in these regimes, rather than an {\it a posteriori} comparison with
numerical results. We explain in more detail the extended validity of the classical perturbation approximation  (CPA) and show that within this approximation, the so-called ``diagonal approximation'' is automatic and does not require ensemble averaging. 
\end{abstract}

\maketitle

The question of stability of quantum motion, originally formulated by
Peres \cite{PERES}, has recently attracted much interest, due to its relevance to quantum computation and decoherence in complex
systems. Peres defined stability in terms of quantum
fidelity $M(t)$, the overlap at time $t$ of two states, which were
identical at time $t=0$, but afterwards propagated in slightly
different dynamical systems, described by Hamiltonians $H^{0}$ and
$H^{V}=H^{0}+V$,  
\begin{equation} 
M(t)=\left\vert \left\langle \psi \left\vert \exp
        \left( iH^{V}t\right) \exp \left( -iH^{0}t\right) \right\vert
      \psi \right\rangle \right\vert ^{2}.
 \label{fidelity} 
\end{equation}  
This quantity is also called Loschmidt echo, because it can
be interpreted as an overlap of a state propagated forward for
time $t$ with $H^{0}$ and then backward for time $t$ with
$H^{V}$, with the initial state. We consider $H^{0}$ to
be strongly chaotic, although our method is not
limited to this case. Even with this restriction, the decay of
fidelity has a surprisingly rich behavior: Most surprising recently
was  the derivation in Ref.~\cite{JALABERT} that for certain range
of perturbations the decay rate is independent of the perturbation
strength. 

The  Loschmidt echo is physically realizable, for example in NMR spin
echo experiments, where back-propagation under a slightly different
Hamiltonian is feasible \cite{NMR_ECHO1,NMR_ECHO2,NMR_ECHO3}.  
There are other examples, which often go unnoticed. An example is
neutron scattering, where the scattering kernel can be written as in
Eq.~(\ref{fidelity}), with $H^{V}$  a momentum boosted version of
$H^{0}$. Many numerical investigations of FD have been
undertaken in various systems
\cite{JACQUOD,CERRUTI,CUCCHIETTI,CUCCHIETTI2,CUCCHIETTI3,CUCCHIETTI4,uniformCERRUTI,PROSEN1,PROSEN2,PROSEN3,ZDINARIC,JACQUOD2,JACQUOD3,SILVESTROV,
BENENTI1,BENENTI2,ECKHARDT,WISNIACKI1,WISNIACKI2,WISNIACKI3,KOTTOS,EMERSON,WEINSTEIN,ADAMOV,WANG,GEORGEOT,VANICEK}.
Depending on the strength of perturbation, there exist at least four qualitatively
different regimes of the decay in chaotic systems \cite{CUCCHIETTI}: As the
perturbation increases, these regimes are perturbative (PT),
Fermi-golden-rule (FGR), Lyapunov (L), and the strong SC regime. 

In the PT regime, in which the characteristic matrix element
of the perturbation is smaller than the mean level spacing $\Delta $,
the decay can be described by a combination of perturbation theory and
random-matrix theory (RMT), and is Gaussian
\cite{CERRUTI,CUCCHIETTI},
\begin{equation} 
M_{PT}\left( t\right) \approx \exp \left( -\overline{V^{2}} \,t^{2}/\hbar ^{2}\right) .  
\label{PT_decay} 
\end{equation}
For intermediate perturbation strengths, the decay follows the Fermi golden rule \cite{JACQUOD} and is exponential,
\begin{equation} 
M_{FGR}\left( t\right) \approx \exp \left( -\Gamma t/\hbar \right)  
\label{FGR_decay} 
\end{equation}
where $\Gamma =2\pi \overline{V^{2}}/\Delta $. In Ref.~\cite{CUCCHIETTI} it is 
shown that this FGR decay is equivalent to the exponential decay
derived semiclassically in Refs.~\cite{JALABERT,CERRUTI}. In other words,
$\Gamma =2K/\hbar $ where $K$ is the classical action diffusion
constant,
\[ 
K=\int_{0}^{\infty }dt\,\left\langle V\left[ \mathbf{r}\left( t\right)
  \right] V\left[ \mathbf{r}\left( 0\right) \right] \right\rangle . 
\]

In the Lyapunov regime, derived in Ref.~\cite{JALABERT}, FD actually does not depend on the strength of
perturbation, but only on the Lyapunov exponent $\lambda$ of the chaotic system,
\begin{equation}
M_{L}\left( t\right) \sim \exp \left( -\lambda t\right) .  
\label{L_decay} 
\end{equation}

We are able to find a numerically feasible {\it uniform} \cite{VANICEK,VANICEK1,VANICEK2,VANICEK3} SC method to 
evaluate FD in the FGR and Lyapunov regimes.  As a result,  we can directly test all approximations made in the derivation of
results (\ref{FGR_decay}) and (\ref{L_decay}) from
Refs. \cite{JALABERT,CERRUTI}. The method starts with a
SC approach based on the CPA \cite{JALABERT,CERRUTI},
and ends with a form of initial value representation (IVR) \cite{ivr,IVR2} which
makes the numerical calculation manageable and the SC
approximation itself more accurate.  

Following notation of
Ref.~\cite{JALABERT}, we want to find  FD for 
an initial Gaussian wave packet  
\[ 
\psi \left( \mathbf{r};0\right) =\left( \pi \sigma ^{2}\right)
^{-d/4}\exp  \left[ \frac{i}{\hbar} \mathbf{p}_{0}\cdot \left
    ( \mathbf{r}-\mathbf{r}_{0}\right)  - \frac{\left
    ( \mathbf{r}-\mathbf{r}_{0}\right) ^{2}}{ 2\sigma ^{2}}\right]. 
\]
It is centered at $\mathbf{r}_{0}$ with dispersion $\sigma $ and has an average 
momentum $\mathbf{p}_{0}.$ We propagate this state with a SC Van Vleck-Gutzwiller 
propagator \cite{vanVleck}
\[ 
K^{sc}(\mathbf{r}^{\prime \prime
  },\mathbf{r}^{\prime };t) = \sum_{j}\left( 2\pi i\hbar \right)
^{-d/2}C_{j}^{1/2}\exp \left( \frac{i}{\hbar} S_{j} - i \frac{\pi}{2}
    \nu _{j} \right).  
\]
Here 
$C_{j}=\left\vert \det \left( \partial
      ^{2}S_{j} / \partial \mathbf{r}^{\prime \prime }\partial
      \mathbf{r}^{\prime }  \right) \right\vert $ 
is the absolute
value of the Van Vleck determinant, $  S_{j}\left( \mathbf{r}^{\prime
    \prime },\mathbf{r}^{\prime };t\right) $ is the action along the
$j$th  trajectory connecting $\mathbf{r}^{\prime}$ with $\mathbf{r}^{\prime \prime }$,   
\[ 
S_{j}\left( \mathbf{r}^{\prime \prime },\mathbf{r}^{\prime
      };t\right) =\int_{0}^{t}dt^{\prime }L\left[ \mathbf{r}\left
      ( t^{\prime }\right) \mathbf{  ,\dot{r}}\left( t^{\prime
      }\right) \mathbf{,}t^{\prime }\right]  
\]  
and $\nu _{j}$ is the Maslov index. 

Expanding each contribution about a central trajectory \cite{TOMSOVIC}, the overlap amplitude of the semiclassically propagated states becomes  \cite{JALABERT} 
\begin{align} 
\label{old_overlap}
&O(t) =\left\langle \psi ^{sc,V}\left( t\right) |\psi ^{sc}\left
    ( t\right) \right\rangle =\left( \sigma ^{2}/\pi \hbar ^{2}\right)
^{d/2}   \int d^{d}r \\ 
&\times \sum_{j,j^{\prime }} \left( C_{j}^{V}C_{j^{\prime
    }}\right) ^{1/2}  \!
 \exp \left[ \frac{i}{\hbar }\left
    ( S_{j}^{V}-S_{j^{\prime }}\right) -\frac{i\pi }{2}\left( \nu
    _{j}^{V}-\nu _{j^{\prime }}\right)   \right] \nonumber \\ 
&\times
    \exp \left\{ -\left
    [ \left( \mathbf{p}_{j}^{\prime }-\mathbf{p}  _{0}\right)
    ^{2}+\left( \mathbf{p}_{j^{\prime }}^{\prime }-\mathbf{p}
    _{0}\right) ^{2}\right] \sigma ^{2}/2\hbar ^{2}\right\}  \nonumber
\end{align}
where $S_j = S_{j}\left( \mathbf{r},\mathbf{r}_{0};t\right)$ and superscript $V$ denotes quantities in the perturbed system.
At this point, two crucial approximations are made in
Refs. \cite{JALABERT, CERRUTI}:  First, only the diagonal terms
  $j=j^{\prime }$ are considered. Ref.~\cite{JALABERT} claims that these are the only terms surviving the average over
  impurities in disordered systems.  Below we show that this is not a separate approximation, but that it follows from the  CPA and does not require any ensemble averaging. 
CPA, the second approximation used in Refs.~\cite{JALABERT,CERRUTI} is based on
an apparently hopeless  assumption that the perturbation does not affect trajectories (i.e., $C_{j}^{V}\approx C_{j}$ and $\nu_{j}^{V}\approx \nu_{j}$)
but only affects the actions, through  
\begin{equation} 
\Delta S_{j}=S_{j}^{V}-S_{j}=-\int_{0}^{t}dt^{\prime }\,V\left
  [ \mathbf{r}  _{j}\left( t^{\prime }\right) \right] .  
\label{action_difference} 
\end{equation}  

Of course this assumption is wrong for individual trajectories which deviate exponentially with time. The reason why the approximation works in quantum mechanics is subtle: The first step to understanding why it yields accurate wave functions lies in the structural stability of the manifolds, as pointed out in Ref.~\cite{CERRUTI}. Assuming that perturbation does not cause a bifurcation and does not significantly change the stable manifold, the evolved manifolds almost exactly overlap whereas the same initial points deviate exponentially by sliding along the manifold \cite{CERRUTI}.

The second step goes as follows: consider trajectories $A(t)$, $A^V(t)$ under the flow $H^0$, $H^V$, respectively. Let $A(0) = A^V(0)$ be a point on the Lagrangian manifold supporting the wave function at $t=0$. While $A^{V}(t)$ exponentially diverges from $A(t)$, if the evolved manifolds (almost) exactly overlap, we can find a point $B(0)$ on the manifold at $t=0$ such that $B^V(t)$ (almost) coincides with $A(t)$. Because of the exponential sensitivity to the initial conditions, point $B(0)$ will be exponentially close to $A(0)$. Trajectories $A(t)$ and $B(t)$ remain exponentially close for all times, so if we use these particular trajectories to find $\psi(t)$ and $\psi^V(t)$, respectively, the CPA will be justified.

 The diagonal approximation and CPA 
  enormously simplify expression
(\ref{old_overlap}) for the overlap amplitude: 
\begin{align} 
\label{position_overlap} 
&O(t)=\left( \sigma ^{2}/\pi \hbar ^{2}\right) ^{d/2} \\ 
&\times \int
d^{d}r\sum_{j}C_{j}\exp \left[ i \Delta S_{j} / \hbar -\left
    ( \mathbf{p}  _{j}^{\prime }-\mathbf{p}_{0}\right) ^{2}\sigma
  ^{2}/\hbar ^{2}\right] . \nonumber 
\end{align}  
At this point, both Refs.~\cite{JALABERT, CERRUTI} resort to 
statistical arguments  to obtain an analytical result. 
Expression~(\ref{position_overlap}) for the overlap would be very
difficult to implement numerically for three reasons: First, in chaotic systems
there is an exponentially growing number of contributing
trajectories. Second, the accuracy would be compromised by proliferating caustic singularities in the Van Vleck determinant $C_{j}$ whenever $\partial \mathbf{r/\partial p}  _{j}^{\prime
  }=0.$ Finally, for each trajectory we would have to perform a
computationally expensive root-search to find initial
$\mathbf{p}_{j}^{\prime }$ that satisfies $  \mathbf{r}\left
  ( \mathbf{r}_{0},\mathbf{p}_{j}^{\prime },t\right) =\mathbf{r} $.
However, there exists a beautiful and simple way to eliminate the
exponential number of contributions, caustic singularities, and the
root search, all at the same time. All three problems can be solved if
we evaluate the overlap (\ref{old_overlap}) in the initial momentum instead of final position representation. Exactly one point on the evolved manifold corresponds to each initial momentum, so no summation is necessary. The new ``Van Vleck determinant'' is exactly 1, so there will be no Maslov indices either. With all these simplifications,
the SC evaluation becomes tractable; in principle, it yields the same result that an arduous evaluation of (\ref{position_overlap}) would:
\begin{align} 
\label{overlap_amplitude} 
&O(t)=\left( \sigma ^{2}/\pi \hbar ^{2}\right) ^{d/2} \int 
d^{d}p^{\prime } \\ 
&\times  \exp \left[ i \Delta S\left[ \mathbf{r}\left( \mathbf{r}_{0},  \mathbf{p}^{\prime },t\right)
    ,\mathbf{r}_{0},t\right] / \hbar  -\left( \mathbf{p}  ^{\prime
      }-\mathbf{p}_{0}\right) ^{2}\sigma ^{2}/\hbar ^{2}\right]. \nonumber 
\end{align}  
The only assumption required to derive (\ref{overlap_amplitude}) is the validity of CPA, in the extended sense described above. Ensemble averaging used in Ref.~\cite{JALABERT} is unnecessary: result (\ref{overlap_amplitude}) works for pure states.   Expression (\ref{overlap_amplitude}) is a special form of IVR
\cite{ivr,IVR2}.  In general, IVR avoids the
singularities and the root search, but at a cost of replacing a sum
over classically allowed paths by an integral over all initial
momenta. In our case, it is even better, since we also eliminated the
integral over final position $\mathbf{r}$.   We remark that
(\ref{overlap_amplitude}) can also be obtained by changing the
integration variable in (\ref{position_overlap}) from final
$\mathbf{r}$ to initial $\mathbf{p}^{\prime }$, but our derivation
avoids the intermediate step (\ref{position_overlap}) that requires making diagonal
approximation in (\ref{old_overlap}). We note the unique property of IVR: in this representation, FD is only due to dephasing. In other representations, the decay can also have a component due to the decay of classical overlaps.

We chose to test our method on the standard
map used in Ref.~\cite{CERRUTI},  
\begin{align*} 
q_{j+1} &=q_{j}+p_{j}~~~(\rm{mod} \ 1), \\ 
p_{j+1} &=p_{j}-\frac{k }{2\pi }\sin \left( 2\pi q_{j+1}\right) ~~~
(  \rm{mod} \ 1). 
\end{align*}   
Perturbation is effected by replacing the parameter $k$ by $k + \epsilon$.  Choice of an $n$-dimensional Hilbert space for the quantized map fixes the effective Planck constant to be $\hbar =\left( 2\pi n\right) ^{-1}$. We note that results of exact quantum and SC computations which we present below are for initial position eigenstate with $q_0 = 0.5$ rather than a wavepacket.
  
\begin{figure}[htbp]
\centerline{\epsfig{figure=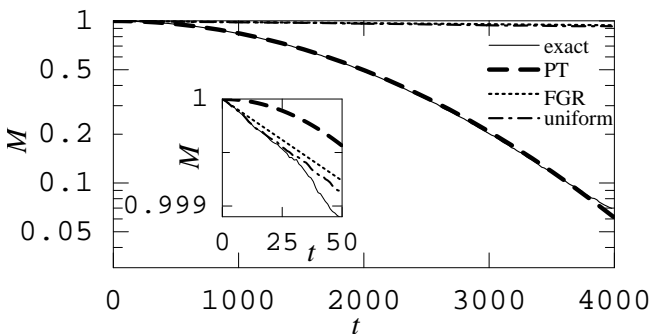,width=\hsize}}
\caption{\label{fidPT} Fidelity in the perturbative regime ($k = 18$, $\lambda \approx 2.21$, $\epsilon = 10^{-4}$, $t_H \approx n = 350$). Inset:  detail for short times.} 
\end{figure}

In previous numerical experiments analytical
predictions of Gaussian or exponential decay have been compared to an
exact quantum calculation: see, e.g., Refs.
\cite{JACQUOD,CERRUTI,CUCCHIETTI,BENENTI1,SILVESTROV}.
While we also have an exact quantum benchmark (FFT) with which to
compare the expressions for various regimes, we reiterate that it
would be hard from a mere comparison of final results for $  M(t)$ to
determine the source of errors. We proceed by discussing how the
uniform method helps to analyze various regimes of decay.  In the
PT regime (see Fig.~\ref{fidPT}), we do not expect any SC approach to
work very well except for short times (much shorter than the
Heisenberg time  $t_{H}=h/\Delta $). The RMT analytical result $M_{PT}$
from Ref. \cite{CERRUTI} gives an excellent agreement in this case. The inset shows, however, that before the Gaussian decay $M_{PT}$ sets in at the Heisenberg time, the uniform expression follows $M_{exact}$ much better. 

   As the perturbation $\epsilon$  increases, we enter the
regimes with exponential decay of fidelity. If the perturbation is strong quantum mechanically, but does not significantly change the stable manifold, CPA may be used.
Even
within the CPA, there are two types of
decay, discussed already in Ref. \cite{JALABERT}. First, there is  decay related
to dephasing of trajectories with uncorrelated actions. Second, there is 
decay related to dephasing of very near trajectories with correlated
actions. For smaller perturbations, the first type of decay is slower and
dominates the behavior of fidelity: this happens in the FGR regime. For
larger perturbations, dephasing of uncorrelated trajectories is so fast
that the quantum overlap is determined by the fraction of near trajectories
that have remained in phase. This is the case in the Lyapunov regime. Transition from the PT to FGR regime occurs for $\epsilon^2 \approx 32 \pi^2 n^{-3} [1 + 2 J_2(k)]^{-1}$ \cite{CERRUTI} when most of the overlap has decayed before Heisenberg time.  Transition from the FGR to Lyapunov regime occurs for $\epsilon^2 \approx 8 \pi^2 \lambda n^{-2} [1 + 2 J_2(k)]^{-1}$ when the FGR decay rate is larger than $\lambda$.

\begin{figure}[htbp]
\centerline{\epsfig{figure=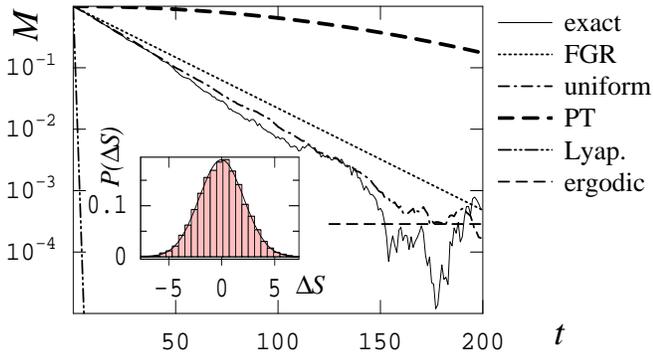,width=\hsize}}
\caption{\label{fidFGR} Fidelity in the FGR regime ($k = 18$, $\lambda \approx 2.21$, $\epsilon = 5 \times 10^{-4}$, $n = 3500$). Horizontal dashed line (``ergodic'') is the limit of FD due to the finite size of Hilbert space. Inset: Histogram of action differences compared to a Gaussian fit.} 
\end{figure}

Using Eq.~(\ref{overlap_amplitude}), fidelity can be written as a weighted average of terms $\exp 
\left[ i(\Delta S^{\prime }\mathbf{-}\Delta S^{\prime \prime })/\hbar \right]
$,
\begin{align}
&M_{unif}\left( t\right) = \left( \frac{\sigma ^{2}}{\pi \hbar ^{2}}\right)
^{d}\int \!
d^{d}p^{\prime } \! \int \! d^{d}p^{\prime \prime }\exp \left[ \frac{i}{\hbar}
(\Delta S^{\prime }\mathbf{-}\Delta S^{\prime \prime }) \right]
\nonumber \\ 
&\times \exp \left\{ -
\left[ \left( \mathbf{p}^{\prime }-\mathbf{p}_{0}\right) ^{2}+\left( \mathbf{
p}^{\prime \prime }-\mathbf{p}_{0}\right) ^{2}\right] \sigma ^{2}/\hbar
^{2}\right\} 
\label{overlap}
\end{align}
where $\Delta S^{\prime \prime }$ corresponds to a trajectory with
initial momentum $\mathbf{p}^{\prime \prime }$. Assuming the averaging window (i.e. the momentum width of the wavepacket) is
large enough, we can make the replacement 
\begin{equation}
\exp \left[ {i (\Delta S^{\prime }\mathbf{-}\Delta
S^{\prime \prime }) / \hbar} \right] 
\approx
\left\langle \exp \left[ {i (\Delta S^{\prime }\mathbf{-}\Delta
S^{\prime \prime }) / \hbar} \right] \right\rangle 
\label{averaging}
\end{equation}
in Eq.~(\ref{overlap}) 
where averaging is over all initial momenta $p^{\prime }$, $p^{\prime \prime
}$. In the FGR regime where dephasing is determined by uncorrelated
trajectories, a further simplification
\begin{equation}
\left\langle \exp \left[ {i (\Delta S^{\prime }\mathbf{-}\Delta
S^{\prime \prime }) / \hbar} \right] \right\rangle \approx \left\langle e^{i
\Delta S^{\prime } / \hbar} \right\rangle \left\langle e^{ -
i \Delta S^{\prime \prime } / \hbar} \right\rangle 
\label{uncorrelated}
\end{equation}
is possible. Due to the central limit theorem, in chaotic systems
distribution of $\Delta S$ approaches a Gaussian and
\begin{equation}
\left\langle \exp \left( i \Delta S / \hbar \right) \right\rangle =\exp 
\left[ i \left\langle \Delta S \right\rangle / \hbar - \sigma
_{\Delta S}^{2} / 2 \hbar ^{2} \right]   \label{Gaussian_average}
\end{equation}
where $\sigma _{\Delta S}^{2}=2Kt$ is the action variance at time $t$.
Applying approximations (\ref{averaging}), (\ref{uncorrelated}), (\ref
{Gaussian_average}) in Eq. (\ref{overlap}), confirms Eq. (\ref{FGR_decay}) for the
FGR decay \cite{JALABERT,CERRUTI}. Fig.~\ref{fidFGR} shows FD in the
FGR regime. In the inset,
histogram of action differences is compared with a Gaussian fit, confirming assumption
(\ref{Gaussian_average}). It is apparent that $M_{unif}$ matches $M_{exact}$
better than the $M_{FGR}$ since $M_{unif}$ takes into account the precise
initial conditions without the averaging assumption (\ref{averaging}) and since $M_{FGR}$ uses an analytic result for $K$, which is only approximate \cite{CERRUTI}. Careful inspection of the short time regime (not shown) reveals that $M_{unif}$ agrees with $M_{exact}$, since unlike $M_{FGR}$, $M_{unif}$ does not depend on the central limit theorem which guarantees the Gaussian assumption (\ref{Gaussian_average}) at later times. Finally, we would like to point out that the uniform expression is very accurate at time $t \approx 120$ when there are approximately $10^{70}$ semiclassical contributions in the sum (\ref{position_overlap})!

\begin{figure}[htbp]
\centerline{\epsfig{figure=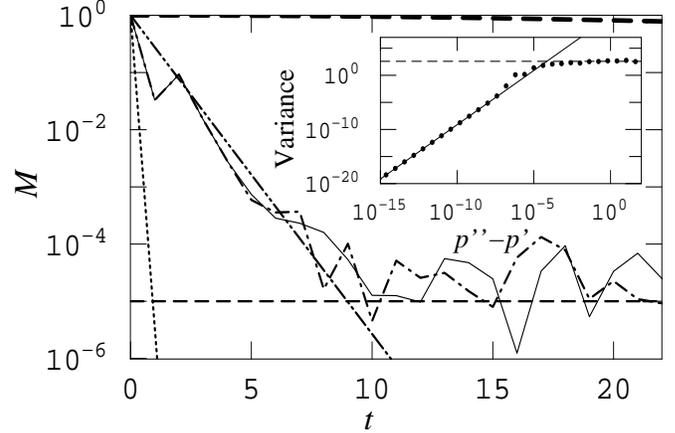,width=\hsize}}
\caption{\label{fidL} Fidelity in the Lyapunov regime ($k = 7$, $\lambda \approx 1.28$, $\epsilon = 5 \times 10^{-4}$, $n = 10^5$). Meaning of lines same as in Fig.~\ref{fidFGR}. Inset: Variance of $\Delta S\left( p^{\prime \prime
}\right) -\Delta S\left( p^{\prime }\right)$ as a function of $p''  - p'$ at time $t=7$. Dots are numerically calculated, dashed line is the horizontal asymptote $2 \sigma_{\Delta S}^2$, solid line is a linear fit for small $p'' - p''$, in agreement with Eq.~(\ref{correlated_variance}).}
\end{figure}

\begin{figure}[htbp]
\centerline{\epsfig{figure=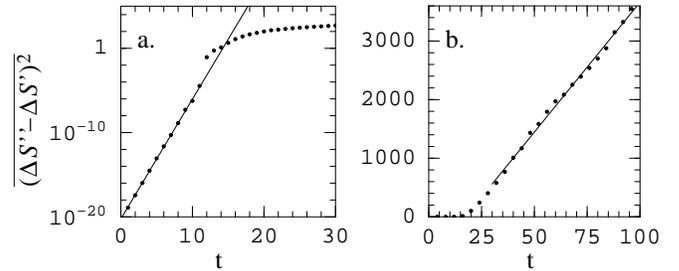,width=\hsize}}
\caption{\label{time_dependence} Variance of $\Delta S\left( p^{\prime \prime
}\right) -\Delta S\left( p^{\prime }\right)$ as a function of $t$ for $p''  - p' = 10^{-11}$: a) exponential dependence for short times
, b) linear dependence for long times
.}
\end{figure}

In the Lyapunov regime, FD is determined by dephasing of
near trajectories with correlated actions \cite{JALABERT}, invalidating
simplification (\ref{uncorrelated}). Now the action difference $\Delta
S^{\prime }\mathbf{-}\Delta S^{\prime \prime }$ depends on the initial
momenta $p^{\prime }$, $p^{\prime \prime }$. Using reasoning similar to Ref.
\cite{JALABERT} or statistical arguments for a random walk with an
exponentially increasing time step \cite{RANDOM_WALK},
it can be shown that
the action difference is also Gaussian distributed, with zero
average and variance
\begin{align}
\left\langle \left[ \Delta S\left( p^{\prime }\right) -\Delta S\left(
p^{\prime \prime }\right) \right] ^{2}\right\rangle  &\approx  ( D /
2\lambda ) e^{2\lambda t}\left( p^{\prime }-p^{\prime \prime }\right) ^{2},
\label{correlated_variance} \\
D &=2\int_{0}^{\infty }dt\,\left\langle V^{\prime }\left( 0\right)
V^{\prime }\left( t\right) \right\rangle .  \nonumber
\end{align}
We can therefore make the replacement
\[
\left\langle \exp \left[ \frac{i}{\hbar }(\Delta S^{\prime }\mathbf{-}\Delta
S^{\prime \prime })\right] \right\rangle \approx \exp \left[ -\frac{D}{4\lambda
\hbar ^{2}}e^{2\lambda t}\left( p^{\prime }-p^{\prime \prime }\right) ^{2}
\right] 
\]
in Eq. (\ref{overlap}) and (\ref{averaging}) to find
\[
M_{L}\left( t\right) \approx \left(1 +  
e^{2\lambda t} D / 2 \lambda \sigma ^{2} \right) ^{-1/2}\approx \left( 2 \lambda \sigma ^{2} / D 
\right) ^{1/2}e^{-\lambda t},
\]
confirming Eq.~(\ref{L_decay}). For the precise definition of $\lambda$, see Ref.~\cite{SILVESTROV}(one has to be careful about the averaging process). Fig.~\ref{fidL} displays $M\left( t\right) $ in the
Lyapunov regime. It shows that
while $M_{L}$ gives an accurate average decay only for $\lambda t\gg 1$, $M_{unif}$ correctly
follows the behavior of $M_{exact}$ even for short times $t \sim \lambda ^{-1}$. The inset shows the variance of $\Delta S\left( p^{\prime
}\right) -\Delta S\left( p^{\prime \prime }\right) $ as a function of
$p^{\prime }-p^{\prime \prime }$ at a fixed time and justifies the assumption made in Ref. \cite{JALABERT}
in derivation of perturbation independent decay: for near trajectories, the variance grows quadratically with $p^{\prime }-p^{\prime \prime }$ (fitted line gives an exponent 2.003), in accordance with Eq.~(\ref{correlated_variance}), while for distant trajectories, in accordance with the derivation of the FGR regime, the variance is independent of $p^{\prime }-p^{\prime \prime }$, 
\begin{equation}
\label{uncorrelated_variance}
\left\langle \left[ \Delta S\left( p^{\prime }\right) -\Delta S\left(
p^{\prime \prime }\right) \right] ^{2}\right\rangle = 2 \sigma _{\Delta S}^{2} = 4 K t.
\end{equation} 
The time dependence of $\left\langle \left[ \Delta S\left( p^{\prime }\right) -\Delta S\left(p^{\prime \prime }\right) \right] ^{2}\right\rangle$ for fixed $p'-p''$ is shown in Fig.~\ref{time_dependence}. Part a) shows that for short times when trajectories are still correlated, this dependence is exponential, in agreement with Eq.~(\ref{correlated_variance}). Part b) shows that for longer times, when correlation is lost, the dependence is linear, as expected from Eq.~(\ref{uncorrelated_variance}).  

To conclude, we have explicitly evaluated SC expressions  which were thought to be intractable numerically, yielding remarkably accurate results for FD in the FGR and Lyapunov regimes. We provided a more detailed explanation why CPA works and employed our method to test other approximations used in Refs.~\cite{JALABERT,CERRUTI}.

This research was supported by the National Science Foundation under Grant No. NSF-CHE-0073544, by ITAMP at the Harvard-Smithsonian Center for Astrophysics, Harvard University, and by the Mathematical Sciences Research Institute at Berkeley. We would like to acknowledge helpful discussions with S. Tomsovic. E.J.H. acknowledges the hospitality of the Max Planck Institute for 
Complex Systems, Dresden and the Humboldt Foundation for hospitality and 
support.

\end{document}